\documentclass{pspum-l}

\usepackage{amssymb}

\usepackage{graphicx}

\usepackage{tikz}                                       %
\usepackage{tikz-cd}                                    %
\usetikzlibrary{cd}                                     %
\usetikzlibrary{calc}                                   %
\usepackage{lipsum}
\usepackage{mathrsfs}

\newtheorem{theorem}{Theorem}[section]

\theoremstyle{definition}

\theoremstyle{remark}

\numberwithin{equation}{section}

\begin{document}

\title[VW Floer Homology, Langlands Correspondence and Categories]{Vafa-Witten Theory: Invariants, Floer Homologies, \\
Higgs Bundles, a Geometric Langlands Correspondence,\\
and Categorification}




\author{Meng-Chwan~Tan}
\address{Department of Physics, National University of Singapore}
\curraddr{}
\email{mctan@nus.edu.sg}
\thanks{This proceeding is based on joint work with Z.-C.~Ong in \cite{OT22}. I would like to thank the referee for questions which have led to further refinement of this proceeding. This proceeding is supported in part by the MOE AcRF Tier 1 grant R-144-000-470-114.}


\date{}

\begin{abstract}
We revisit Vafa-Witten theory in the more general setting whereby the underlying moduli space is not that of instantons, but of the full Vafa-Witten equations. We physically derive (i) a novel Vafa-Witten four-manifold invariant associated with this moduli space, (ii) their relation to Gromov-Witten invariants, (iii) a novel Vafa-Witten Floer homology assigned to three-manifold boundaries, (iv) a novel Vafa-Witten Atiyah-Floer correspondence, (v) a proof and generalization of a conjecture by Abouzaid-Manolescu 
 about the hypercohomology of a perverse sheaf of vanishing cycles, (vi) a Langlands duality of these invariants, Floer homologies and hypercohomology, and (vii) a quantum geometric Langlands correspondence with purely imaginary parameter that specializes to the classical correspondence in the zero-coupling limit, where Higgs bundles feature in (ii), (iv), (vi) and (vii). We also explain how these invariants and homologies will be categorified in the process, and discuss their higher categorification. We thereby relate differential and enumerative geometry, topology and geometric representation theory in mathematics, via a maximally-supersymmetric topological quantum field theory with electric-magnetic duality in physics.
\end{abstract}

\maketitle

\section{Vafa-Witten Twist of $\mathcal{N}=4$ Gauge Theory, and a Vafa-Witten Invariant}\label{vwgeneral}
In this section, we start by reviewing aspects of Vafa-Witten (VW) theory on a four-manifold $M_4$ with a  real, simple and compact gauge group $G$ necessary for this work, referring to ~\cite{VW94, LL97}.

\subsection{Vafa-Witten Theory} \label{VW twist discussion}
VW theory is an $\mathcal{N}=4$ topological quantum field theory (TQFT) on a four-manifold $M_4$ with a single scalar supercharge $\mathcal{Q}$ and complexified gauge coupling $\tau$. 
The localization equations, also known as VW equations, are obtained by setting to zero $\{\mathcal{Q},\text{fermion}\}$: 

\begin{equation}\label{4dbps0}
\begin{split}
  F^{+}_{\mu\nu}+\frac{1}{4}[B_{\mu\rho},B_{\lambda\nu}]g^{\rho\lambda}=0,\\
    \mathcal{D}_{\nu}B^{\nu\mu}=0.
\end{split}
\end{equation}
These constitute the BPS equations for the theory, with the zero modes of $A_{\mu}$ and $B_{\mu\nu}$ that satisfy \eqref{4dbps0} defining a moduli space which the path integral localizes on.\footnote{In the full VW equations, there is a bosonic scalar field $C$ that is also a generator of gauge transformations. We thus set the zero mode of $C$ to vanish since we wish to consider only irreducible connections, giving us \eqref{4dbps0} as the resulting BPS equations.}
The bosonic part of the VW action involving only $A$ and $B$ will be of the form
\begin{equation} \label{4dsk}
    \frac{1}{e^2}\int_{M_4} \text{Tr}(|s|^2+ |k|^2) + \text{topological term},
\end{equation}
with
\begin{equation}\label{sandk}
    \begin{aligned}
        s_{\mu\nu} &= F^{+}_{\mu\nu}+\frac{1}{4}\big[B_{\mu\rho}, B^{\rho}_{\nu}\big],\\
        k_{\nu} &= \mathcal{D}^{\mu}B_{\mu\nu}.
    \end{aligned}
\end{equation}
We further note that the 2-form $B$ need not vanish if the scalar curvature of\emph{ K\"ahler} $M_4$ and the gauge group $G$ are not simultaneously non-negative and locally a product of $SU(2)$’s~\cite{VW94}, and we will assume this to be the case here. The topological term in \eqref{4dsk} is 
\begin{equation}\label{4d topological term}
     -\frac{ i \tau }{4\pi}\int_{M_4} \text{Tr}\, \bigg(F\wedge F  
    +  dB\wedge \star DB + B\wedge d(\star DB) \bigg)
\end{equation}
where we have taken the liberty to add a $\mathcal{Q}$-exact term after 
$F\wedge F$ that is null in the spectrum of VW theory given by the $\mathcal {Q}$-cohomology, for later convenience. Also, we have used the fact that $B$ is self-dual, whence $\star B = B$, and here, $D=d+A$ where $\star DB$ is a one-form on $M_4$. 

\subsection{A Vafa-Witten Invariant}
VW theory is a balanced TQFT (same number of fermion pair zero
modes), and the path integral localizes to a virtually zero-dimensional moduli
space $\mathcal{M}_{\text{VW}}$ of solutions to (\eqref{4dbps0}), whence the nonvanishing topological invariant is the partition function that can be interpreted as an integral of a virtual zero-form  on virtually zero-dimensional ${\mathcal M}_{\text{VW}}$:
\begin{equation} \label{vwk}
   \mathcal{Z}_{\text{VW},M_4}(\tau, G) = \sum_{k}  a_k q^{m_k}. 
\end{equation}
Here, $q = e^{2\pi i \tau}$, $k$ denotes the $k^{\text{th}}$ sector of $\mathcal{M}_{\text{VW}}$, the number $a_k$ is given by
\begin{equation} \label{ak-0}
\boxed{a_k =  \int_{\mathcal{M}^k_{\text{VW}}} \Omega^0 \wedge e({T_{\mathcal{M}^k_{\text{VW}}}}), \quad \text{where \, $\Omega^0(\mathcal{M}^k_{\text{VW}})=(1 + B^4)^{dim_{\mathbb{C}}\mathcal{M}^k_{\text{VW}}}$}}
\end{equation}
$B$ is a coordinate on $\mathcal{M}^k_{\text{VW}}(A,B)$, $e$ is the signed Euler class of the tangent bundle ${T_{\mathcal{M}^k_{\text{VW}}}}$, and $m_k$ is the corresponding VW number given by
\begin{equation} \label{m_k}
 \boxed{  m_k =  \frac{ 1}{8\pi^2}\int_{M_4} \text{Tr}\, \bigg(F_{(k)}\wedge F_{(k)}  +  dB_{(k)}\wedge \star DB_{(k)} + B_{(k)}\wedge d(\star DB_{(k)})\bigg) }
\end{equation}
 
Notice that $\mathcal{Z}_{\text{VW},M_4}$ is a topological invariant of $M_4$ which is an algebraic count of VW solutions with corresponding weight given by  $a_k q^{m_k}$ that we elaborated on above. This defines a novel $\tau$-dependent Vafa-Witten invariant of $M_4$.\footnote{A purely algebro-geometric definition of $\mathcal{Z}_{\text{VW},M_4}$, in particular the $a_k$'s, was first given by Tanaka-Thomas in~\cite{TT17}, albeit for projective algebraic surfaces only. The novelty here is that we provide a purely differentio-geometric definition of the $a_k$'s for a more general $M_4$. }

When $B = 0$, $a_k$ will become the Euler characteristic $\chi(\mathcal{M}^k_{\text{inst}})$, while $m_k$ will become the instanton number. Then, $\mathcal{Z}_{\text{VW},M_4}$ will just become the usual partition function for instantons first derived in~\cite{VW94}, as expected.

\section{An $\mathcal{N}=(4,4)$ $A$-model, Higgs Bundles and Gromov-Witten Theory}\label{sigmareduction}
In this section, we will perform dimensional reduction of the 4d VW theory down to 2d. The four-manifold $M_4$ will be taken to be $M_4 = \Sigma \times C$, where $\Sigma$ and $C$ are both closed Riemann surfaces, and $C$ is of genus $g\geq 2$.  

\subsection{Finiteness Conditions, BPS Equations in 2d and an $\mathcal{N} = (4,4)$ Sigma-Model} \label{4d bps to 2d bps}
We consider a block diagonal metric $g$ for $M_4 = \Sigma \times C$, 
\begin{equation}\label{4d metric}
    g = \text{diag}\big(g_{\Sigma}, \epsilon g_C \big),
\end{equation}
where $\epsilon$ is a small parameter to deform $g_C$. We shall use capital letters $A,B=x^1,x^2$ to denote coordinates on $\Sigma$, and small letters $a,b=x^3,x^4$ to denote coordinates on $C$. Taking the limit $\epsilon\to0$ then gives us a 2d theory on $\Sigma$ with $\mathcal{N} = (4,4)$ supersymmetry.

The topological term aside, terms in \eqref{4dsk} with $\mu,\nu,\rho=A,B$ vanish as $\epsilon\to0$, while those with $\mu,\nu,\rho=A,b$ survive. For $\mu,\nu,\rho=a,b$, each term must be set to zero individually as they are accompanied by a factor of $\epsilon^{-1}$. Since the action \eqref{4dsk} is a sum of squares of such terms, we will need to set them to zero. This constraint will  give us the finiteness conditions.

Before we proceed further, we note the fact that $F^{+}_{\mu\nu}=\frac{1}{2}(F_{\mu\nu}+\frac{1}{2}\epsilon_{\mu\nu\rho\lambda}F^{\rho\lambda})$, and that $B_{\mu\nu}$ is an anti-symmetric and self-dual 2-form ($B_{\mu\nu}=\frac{1}{2}\epsilon_{\mu\nu\rho\lambda}B^{\rho\lambda}$) with 3 independent components which we can take to be $B_{12}$, $B_{13}$ and $B_{14}$.

The first finiteness condition we obtain by using the self-duality property of $B_{\mu\nu}$ is
\begin{equation}\label{finite3}
    D_{3}B_{12}=-D_{4}B_{12}=0.
\end{equation}
The field $B_{12}$ is a 0-form w.r.t rotations on both $C$ and $\Sigma$, so \eqref{finite3} tells us that the 0-form $B_{12}$ is covariantly constant on $C$, which means $B_{12}$ generates infinitesimal gauge transformations while leaving $A_C$ fixed. We can however set $B_{12}=0$, since we require gauge connections to be irreducible to avoid complications on $\mathcal{M}^G_H(C)$. 

Next, identifying $B_{13}$ and $B_{14}$ as the two components of a 1-form $\varphi$ on $C$ and using the self-duality properties of $F^+$ and $B_{\mu\nu}$, we obtain Hitchin's equations on $C$~\cite{H87} as the second finiteness condition, given by\footnote{$D^{*}\varphi = \star D \star \varphi = D_{\mu}\varphi^{\mu}$, where $\star$ is the Hodge star operator.}
\begin{equation}\label{Hitchin eqns}
\begin{split}
   F_C - \varphi\wedge\varphi & = 0, \\
  D\varphi = D^{*}\varphi & = 0,
\end{split}
\end{equation}
where
\begin{equation} \label{varphi}
 \varphi = B_{13}dx^3 + B_{14}dx^4 =\varphi_{3}dx^3+\varphi_{4}dx^4. 
\end{equation}
The space of solutions of $(A_C, \varphi)$ to \eqref{Hitchin eqns} modulo gauge transformations then span Hitchin's moduli space $\mathcal{M}^G_H(C)$ for a connection $A_C$ on a principal $G$-bundle $P$ over the Riemann surface $C$, and a section $\varphi\in\Omega^1(C)$. The above equations leave the $(x^1, x^2)$ dependence of $A_C$ and $\varphi$ arbitrary, and thus the fields $(A_C, \varphi)$ define a map $\Phi: \Sigma \to \mathcal{M}^G_H(C)$. 
The target space $\mathcal{M}^G_H(C)$ is a hyper-K\"{a}hler manifold, whence the sigma-model on $\Sigma$ has an $\mathcal{N}=(4,4)$ supersymmetry ~\cite{AF81}. 

To obtain the corresponding 2d BPS equations of the ${\mathcal N} = (4,4)$ sigma model on $\Sigma$, we perform dimensional reduction of \eqref{sandk} on $C$ with $s=k=0$. Noting the fact that only terms with mixed indices on $\Sigma \times C$ survive the reduction on $C$, together with the self-duality properties of $B_{\mu\nu}$, we obtain, from \eqref{sandk} and $s=k=0$,
\begin{equation}\label{effbps}
    \begin{split}
       F_{Aa}^{+} &= 0, \\
        \mathcal{D}_{A}B^{Aa} &=0.
    \end{split}
\end{equation}
Switching to complex coordinates, \eqref{effbps} can be written as $\partial_{\bar{z}}A_{\bar{w}}=\partial_{\bar{z}}\varphi_{w}=0$.\footnote{In complex coordinates, we have $z=x^1+ix^2$ and $w=x^3+ix^4$, where $A_{\bar{w}}=\frac{1}{2}(A_3 + iA_4)$ and $\varphi_{w}=\frac{1}{2}(B_{13}-i B_{14})$.} With $A_{\bar{w}}$ and $\varphi_{w}$ corresponding to bosonic scalars $X^{i}$ and $Y^{i}$ in the sigma-model, respectively, we get the 2d BPS equations as
\begin{equation}\label{2dbpsfull}
    \begin{split}
        \partial_{\bar{z}}X^{i} &= 0,\\
        \partial_{\bar{z}}Y^{i} &= 0.
    \end{split}
\end{equation}
After suitable rescalings, we can then rewrite \eqref{4dsk} (with $idz\wedge d\bar{z}=|dz^2|)$ as 
\begin{equation}\label{2dfinalaction}
\begin{split}
    S_{\text{2d}}&=\frac{1}{e^2}\int_{\Sigma}|dz^2| g_{i\bar{j}}\bigg(\partial_{z}X^{\bar{i}}\partial_{\bar{z}}X^{j}+\partial_{z}X^{i}\partial_{\bar{z}}X^{\bar{j}}+\partial_{z}Y^{\bar{i}}\partial_{\bar{z}}Y^{j}+\partial_{z}Y^{i}\partial_{\bar{z}}Y^{\bar{j}} \bigg)\\&+ \text{topological term}.
\end{split}
\end{equation}
Hence, the path integral of the 2d, ${\mathcal N} = (4,4)$ sigma model on $\Sigma$ with action \eqref{2dfinalaction}, localizes on the moduli space of holomorphic maps $\Phi(X^i, Y^i):\Sigma\to\mathcal{M}^G_H(C)$:
\begin{equation}\label{mmaps2d}
    \mathcal{M}_{\text{maps}} = \{ \Phi(X^i, Y^i):\Sigma\to\mathcal{M}^G_H(C) \;|\; \partial_{\bar{z}}X^{i}=\partial_{\bar{z}}Y^{i}=0 \},
\end{equation}
where we have a 2d ${\mathcal N} = (4,4)$ $A$-model on $\Sigma$ with target $\mathcal{M}^G_H(C)$.
 \subsection{An $A$-model in Complex Structure $I$}
 The space of fields $(A_C$, $\varphi)$ span an infinite-dimensional affine space $\mathcal{W}$. The cotangent vectors $\delta A_C$ and $\delta \varphi$ to $\mathcal{M}^G_H(C)$ are solutions to the variations of equations \eqref{2dbpsfull}. We can then introduce a basis $(\delta A_w, \delta \varphi_{\bar{w}})$ and $(\delta A_{\bar{w}}, \delta \varphi_{w})$ in $\mathcal{W}$.
From the BPS equations \eqref{2dbpsfull}, which are $\partial_{\bar z} A_{\bar w} =0$ and $\partial_{\bar z} \varphi_w = 0$, one can see that the complex structure relevant to the $A$-model is $I$, with linear holomorphic functions consisting of $A_{\bar{w}}$ and $\varphi_w$. In complex structure $I$, $\mathcal{M}^G_H(C)$ can be identified as the moduli space of stable Higgs $G$-bundles on $C$,  $\mathcal{M}^G_{\text{Higgs}}(C)$. One can write the corresponding symplectic form as $\omega_I = \omega'_I-\delta \lambda_I$, where
\begin{equation}\label{omegaexact}
    \omega'_I = -\frac{1}{4\pi}\int_C\,\text{Tr}\,\delta A_C \wedge\delta A_C \quad \text{and} \quad \lambda_I = \frac{1}{4\pi}\int_C\,\text{Tr}\,\varphi\wedge\delta\varphi,
\end{equation}
and $\omega_I$ is cohomologous to $\omega'_I$. 
Comparing the 4d topological term in \eqref{4d topological term} to \eqref{omegaexact}, we see that the topological term can be written as 
\begin{equation}
    i\tau\int_{\Sigma}\,\Phi^{*}(\omega_I).
\end{equation}
The 2d action \eqref{2dfinalaction}, including the topological term, is then 
\begin{equation}\label{2dfinalaction2}
\begin{split}
    S_{\text{2d}}&=\frac{1}{e^2}\int_{\Sigma}|dz^2| g_{i\bar{j}}\bigg(\partial_{z}X^{\bar{i}}\partial_{\bar{z}}X^{j}+\partial_{z}X^{i}\partial_{\bar{z}}X^{\bar{j}}+\partial_{z}Y^{\bar{i}}\partial_{\bar{z}}Y^{j}+\partial_{z}Y^{i}\partial_{\bar{z}}Y^{\bar{j}} \bigg)\\&+i\tau\int_{\Sigma}\Phi^{*}(\omega_I).
\end{split}
\end{equation}
We thus have a 2d, ${\mathcal N} = (4,4)$ $A$-model on $\Sigma$ with target $\mathcal{M}^G_{\text{Higgs}}(C)$, where the path integral localizes on 
\begin{equation} \label{Mmaps final}
    \mathcal{M}_{\text{maps}} = \{ \Phi(X^i, Y^i):\Sigma\to\mathcal{M}^G_{\text{Higgs}}(C) \;|\; \partial_{\bar{z}}X^{i}=\partial_{\bar{z}}Y^{i}=0 \},
\end{equation}
the moduli space of holomorphic maps $\Phi:\Sigma\to \mathcal{M}^G_{\text{Higgs}}(C)$. 

\subsection{Vafa-Witten Invariants as Gromov-Witten Invariants of Higgs Bundles}

The virtual dimension of $\mathcal{M}_{\text{maps}}$, like that of ${\mathcal M}_{\text {VW}}$, ought to also be zero. This is because the 2d $A$-model is obtained via a topological deformation that sets $C\to0$ in the original 4d VW theory, whence the relevant index of kinetic operators counting the dimension of moduli space remains the same.

Like ${\mathcal Z}_{\text{VW}, M_4}$ in 4d, $\mathcal{Z}^{\text{closed}}_{A,\Sigma}$ can be interpreted as an integral of a virtual zero-form on virtually zero-dimensional ${\mathcal M}_{\text {maps}}$, whence it can be evaluated as 
\begin{equation}
   \mathcal{Z}^{\text{closed}}_{A,\Sigma}(\tau, \mathcal{M}^G_{\text{Higgs}}(C)) = \sum_l {\tilde a}_l q^{{\tilde m}_l}.
\end{equation}
Here, $l$ denotes the $l^{\text{th}}$ sector of ${\mathcal M}_{\text {maps}}$  defined in \eqref{Mmaps final} for \emph{genus one} $\Sigma$, the rational number ${\tilde a}_l$ is given by
 \begin{equation} \label{al-0}
 \boxed{{\tilde a}_l = \int_{{\mathcal M}^l_{\text {maps}}} e(\mathcal{V})}
 \end{equation}
where $e$ is the signed Euler class of the vector bundle $\mathcal{V}$ with fiber $H^0(\Sigma, K \otimes \Phi^*T^*{{\mathcal M}^l_{\text {maps}}})$ and canonical bundle $K$ on $\Sigma$, and ${\tilde m}_l$ is the corresponding worldsheet instanton number given by
\begin{equation} \label{q_l-0}
    \boxed{{\tilde m}_l = \frac{1}{2\pi}\int_{\Sigma}\,\Phi^{*}_l(\omega_I)}
\end{equation}

Notice that $\mathcal{Z}^{\text{closed}}_{A,\Sigma}$ is an enumerative invariant which is an algebraic count of holomorphic maps with corresponding weight given by ${\tilde a}_l q^{{\tilde m}_l}$ that we elaborated on above. This coincides with the definition of the GW invariant, which then means that one can identify $\mathcal{Z}^{\text{closed}}_{A,\Sigma}$ as 
\begin{equation}\label{GW invariant}
    \mathcal{Z}_{\text{GW},\Sigma}(\tau, \mathcal{M}^G_{\text{Higgs}}(C)) = \sum_l {\tilde a}_l q^{{\tilde m}_l}
\end{equation}
where $\mathcal{Z}_{\text{GW},\Sigma}$ is a $\tau$-dependent GW invariant of $\mathcal{M}^G_{\text{Higgs}}(C)$.
From the topological invariance of the 4d theory, we have a 4d-2d correspondence of partition functions 
\begin{equation}\label{VW=GW}
    \mathcal{Z}_{\text{VW},M_4}(\tau, G) = \mathcal{Z}_{\text{GW},\Sigma}(\tau, \mathcal{M}^G_{\text{Higgs}}(C)) .
\end{equation}
In other words, we have a correspondence between the VW invariant of $M_4 = \Sigma \times C$ and the GW invariant of $\mathcal{M}^G_{\text{Higgs}}(C))$.

In fact, recall that the numbers ${\tilde m}_l$ (in \eqref{GW invariant}) correspond to the numbers $m_k$ (in \eqref{vwk}). Hence, \eqref{VW=GW} means that we have 
\begin{equation}
\label{ak=al}
 \boxed{   a_k = \tilde{a}_l}
\end{equation}
where $a_k$ and $\tilde{a}_l$ are given in \eqref{ak-0} and \eqref{al-0}, respectively. In other words, one can also determine the $a_k$'s, the VW invariants of $T^2 \times C$, via the signed Euler class of a bundle $\mathcal{V}$ over $\mathcal{M}^l_\text{maps}$.\footnote{Computing the ${\tilde a}_l$'s and thus $a_k$'s for $T^2 \times C$ explicitly is a purely mathematical endeavour that is beyond the scope of this physical mathematics proceeding which main objective is to furnish their fundamental definitions via the expressions \eqref{al-0} and \eqref{ak-0}, respectively. The reader who seeks an explicit computation of these invariants may be happy to know that after our work appeared, this was done purely mathematically in~\cite{N23}.}



\section{A Novel Floer Homology from Boundary Vafa-Witten Theory}\label{vwsqm}
In this section, we will show how we can physically derive a novel Floer homology by considering boundary VW theory on $M_4 = M_3 \times \mathbb{R}^+$ to physically derive a VW Floer homology assigned to $M_3$.\footnote{To be precise, VW theory is still being defined on an $M_4$ with no boundary. However, to make contact with Floer theory, we will need to examine a hyper-slice of $M_4$, which we can topologically regard as $M_3 \times \mathbb{R}^- \cup_{M_3} M_3 \times \mathbb{R}^+$. 
As there is no time-evolution in our topological theory, it is sufficient to examine only $M_3 \times \mathbb{R}^+$, where $M_3$ can then be regarded as a boundary. 
This is consistent with the idea that categorification of topological invariants can be achieved via successive introductions of boundaries to $M_4$, which we will elaborate upon in \S\ref{sec:web of dualities}.}

\subsection{SQM Interpretation of Boundary Vafa-Witten Theory}
Let the manifold of the 4d theory in \eqref{4dsk} be $M_4 = M_3 \times \mathbb{R}^+$, where the $M_3$ boundary is a closed three-manifold, and $\mathbb{R}^+$ is the `time' coordinate. We also let spacetime indices take the values $\mu = 0,1,2,3$, with $\mu=0$ being the time direction, while $\mu = i,j,k = 1,2,3$ being the spatial directions.

Turning to the BPS equations \eqref{4dbps0} of boundary VW theory, we split the indices into space and time directions. Using $F^{+}_{\mu\nu}=\frac{1}{2}(F_{\mu\nu}+\frac{1}{2}\epsilon_{\mu\nu\rho\lambda}F^{\rho\lambda})$ and $B_{\mu\nu}=\frac{1}{2}\epsilon_{\mu\nu\rho\lambda}B^{\rho\lambda}$, we can reexpress the VW equations \eqref{4dbps0} as
\begin{equation}\label{vwflow}
    \begin{aligned}
        \dot{A}^{i}+\frac{1}{2}\epsilon^{ijk}\big(F_{jk}-[B_{j}, B_{k}] \big) &= 0,\\
        \dot{B}^{i} + \epsilon^{ijk}\big(\partial_j B_k + [A_j, B_k]\big) &= 0,
    \end{aligned}
\end{equation}
where the temporal gauge $A^0=0$ is taken, $B^i = B^{0i}$, $\epsilon^{ijk}=\epsilon^{0ijk}$, and $A^i, B^{i}\in \Omega^1(M_3)$.\footnote{Using self-duality properties, we have $B^{0i} = B^i = \epsilon^{ijk}B_{jk}$. }

Introducing a complexified connection $\mathcal{A}=A+iB \in \Omega^1(M_3)$, of a $G_\mathbb{C}$-bundle on $M_3$. We then find that \eqref{vwflow} can be expressed  as 
\begin{equation}\label{flatcomplex}
  {  \dot{\mathcal{A}}^i+\frac{1}{2}\epsilon^{ijk}\mathcal{F}_{jk}=0, }
\end{equation}
where $\mathcal{F}\in \Omega^2(M_3)$ is the complexified field strength. Note that $\epsilon^{ijk}\mathcal{F}_{jk}$ is a gradient vector field of a complex Chern-Simons functional
\begin{equation} \label{complex CS fn}
 V(\mathcal{A})= CS(\mathcal{A}) = -\frac{1}{4\pi^2}\int_{M_3}\text{Tr}\bigg(\mathcal{A}\wedge d\mathcal{A} + \frac{2}{3}\mathcal{A}\wedge \mathcal{A}\wedge \mathcal{A}\bigg), 
\end{equation}
where we have the gradient flow equation
\begin{equation}\label{flatcomplex - final}
\frac{d{\mathcal{A}}^i}{dt}+ s g_{\mathfrak A}^{ij}\frac{\partial V({\mathcal A})}{\partial {\mathcal A}^{j}}=0 .
\end{equation}
The action for boundary VW theory in \eqref{4dsk} can be rewritten as 
\begin{equation}\label{vwsqmcomplex}
\begin{split}
    S_{\text{VW}}^{\text{bdry}} &= \frac{1}{e^2}\int dt \int_{M_3}\text{Tr}\bigg( \dot{\mathcal{A}}^i + s g_{\mathfrak A}^{ij}\frac{\partial V({\mathcal A})}{\partial {\mathcal A}^{j}}\bigg)^2 \\
    &-\frac{i\tau}{4\pi}\int_{M_3}\text{Tr}\bigg(A\wedge dA + \frac{2}{3}A\wedge A\wedge A + B\wedge \star D B\bigg).
\end{split}
\end{equation}

We can interpret \eqref{vwsqmcomplex} as the action of an SQM model with target $\mathfrak A$, the space of complexified connections $\mathcal{A}$ on $M_3$ with metric $g_{\mathfrak A}^{ij}$, and a single nilpotent topological scalar supercharge ${\mathcal Q}$, where \eqref{flatcomplex - final}, which describes the VW equations, can be interpreted as a gradient flow equation between fixed (or time-invariant) critical points of $CS(\mathcal{A}) $ on $\mathfrak A$. Assuming that the fixed critical points are isolated and nondegenerate in $\mathfrak A$,\footnote{This is guaranteed (though not necessary) when all critical points are isolated and nondegenerate.  This can be the case for an appropriate choice of $G$ and $M_3$. For example, one could choose (1) $G$ compact and $M_3$ of nonnegative Ricci curvature such as a three-sphere or its quotient, or (2) an $M_3$ with a finite $G_{\mathbb C}$ representation variety, and introduce physically-trivial $\mathcal Q$-exact terms to the action to perturb $V({\mathcal A})$. We would like to thank A.~Haydys for discussions on this. \label{isolated}} the partition function of boundary VW theory will thus be an algebraic count of fixed critical points of $CS(\mathcal{A})$ , i.e., fixed flat $G_\mathbb{C}$-connections on $M_3$, where there are VW flow lines between these fixed critical points described by the gradient flow equation.




The second term in \eqref{vwsqmcomplex} is a $\tau$-dependent topological term that contributes to an overall factor in the path integral.

\subsection{A Novel Vafa-Witten Floer Homology}
For an $M_4$ with boundary $\partial M_4=M_3$, one needs to specify boundary conditions on $M_3$ to compute the path integral by first defining a restriction of the fields to $M_3$, which we shall denote as $\Psi_{M_3}$, and then specifying boundary values for these restrictions. This is equivalent to inserting in the path integral an operator  functional ${F}(\Psi_{M_3})$ that is nonvanishing in the ${\mathcal Q}$-cohomology. The partition function on $M_4$ can be expressed as 
\begin{equation}\label{4d3dpartition}
    {\mathcal{Z}_{\text{VW},M_4}(\tau, G) = \langle 1 \rangle_{{F}(\Psi_{M_3})} = \sum_k  {\mathcal F}^{G,\tau}_{\text{VW}}(\Psi_{M_3}^k)}.
\end{equation}
Here, the summation in `$k$' is over all sectors of $\mathcal{M}_{\text{VW}}$ labeled by the VW number $m_k$, and 
${\mathcal F}^{G,\tau}_{\text{VW}}(\Psi_{M_3}^k)$ is the $k^{\text{th}}$ contribution to the partition function that depends on the expression of ${{F}(\Psi_{M_3})}$ in the bosonic fields on $M_3$ evaluated over the corresponding solutions of the VW equations restricted to $M_3$.


As an SQM model on $\mathfrak A$, the partition function can be expressed as
\begin{equation} \label{4d3dpartition - 2nd}
     \mathcal{Z}_{\text{VW},M_4}(\tau, G) =   \sum_k  {\mathcal F}^k_{\text{VW-Floer}} (M_3, G, \tau), 
\end{equation}
where each ${\mathcal F}^k_{\text{VW-Floer}} (M_3, G, \tau)$ can be identified with a class in what we shall henceforth call a Vafa-Witten Floer homology $\text{HF}^{\text{VW}}_{d_k}(M_3, G, \tau)$ assigned to $M_3$ of degree $d_k$, where $d_k$ counts the number of outgoing VW flow lines from the corresponding fixed critical point of $CS(\mathcal A)$.

In summary, from \eqref{4d3dpartition} and \eqref{4d3dpartition - 2nd}, we can write 
\begin{equation}\label{4d3dpartitionfinal}
    \mathcal{Z}_{\text{VW},M_4}(\tau, G)  = \sum_k {\mathcal F}^{G,\tau}_{\text{VW}}(\Psi_{M_3}^k)
    =\sum_k \text{HF}_{d_k}^{\text{VW}}(M_3, G, \tau) = \mathcal{Z}^{\text{Floer}}_{\text{VW},M_3}(\tau,G)
\end{equation}
where `$k$' sums from zero to the maximum number of fixed VW solutions on $M_3 \times \mathbb{R}^+$ that correspond to isolated and non-degenerate fixed critical points of $CS(\mathcal A)$.\footnote{See footnote~\ref{isolated}.}

The $\tau$-dependence of ${\mathcal F}^{G,\tau}_{\text{VW}}(\Psi_{M_3}^k)$ and therefore $\text{HF}_k^{\text{VW}}(M_3, G, \tau)$ arises because in evaluating \eqref{4d3dpartition}, there will be a factor of $q^{s_k}$ for the $k^{\text{th}}$ term, where from the action $S^{\text{bdry}}_{\text{VW}}$ in \eqref{vwsqmcomplex}, the  number   
\begin{equation} \label{sk}
    s_k = \frac{1}{8\pi^2}\int_{M_3} \text{Tr}\bigg(A_{(k)}\wedge dA_{(k)} + \frac{2}{3}A_{(k)}\wedge A_{(k)}\wedge A_{(k)} + B_{(k)}\wedge \star D B_{(k)}\bigg).
\end{equation}
Here, the subscript `$(k)$' denotes that they are the $k^{\text{th}}$ fixed solution to the VW equations on $M_3 \times \mathbb{R}^+$ restricted to $M_3$.

\section{A Vafa-Witten Atiyah-Floer Correspondence}\label{vwafconj}
\begin{figure}
    \centering
    \includegraphics[width=0.6\linewidth]{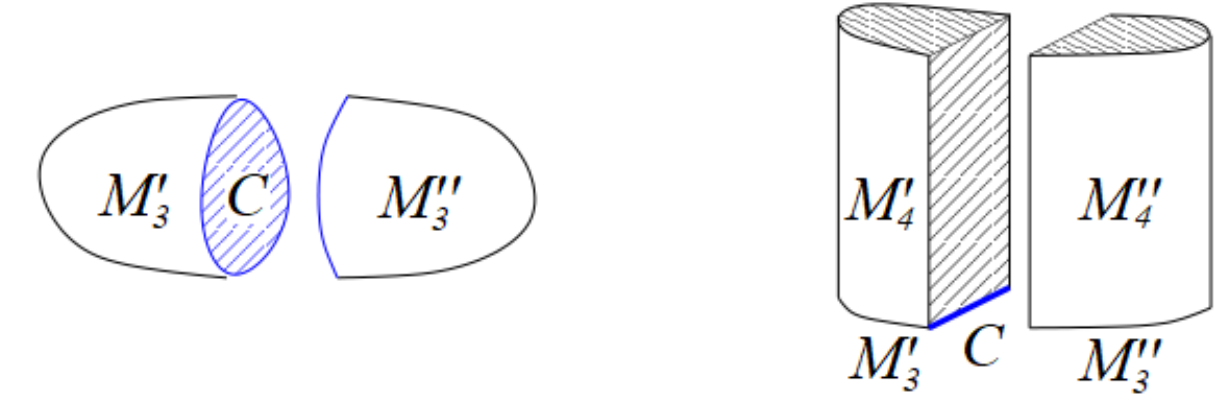}
    \caption{\textbf{Left:} $M_3$ as a connected sum of three-manifolds $M'_3$ and $M''_3$ along a common Riemann surface $C$. \textbf{Right:} $M_4$ split along four-manifolds $M'_4$ and $M''_4$ with corners. }
    \label{fig:heegaardsplit}
\end{figure}

In this section, we continue with a Heegaard split of $M_3$ into $M'_3$ and $M''_3$ along a Riemann surface $C$, as shown in Fig.~\ref{fig:heegaardsplit} (left), allowing us to relate Vafa-Witten Floer homology obtained in the previous section to Lagrangian Floer homology, in what is a novel Vafa-Witten version of the Atiyah-Floer correspondence~\cite{A81} based on instantons. In doing so, we would be able to physically prove and generalize a conjecture by mathematicians Abouzaid-Manolescu about the hypercohomology of a perverse sheaf of vanishing cycles in the moduli space of irreducible flat $SL(2, \mathbb{C})$-connections on $M_3$.  
 We can thus write $M_4=\big(\mathbb{R}^+ \times I'\times C\big) \cup_{C} \big(\mathbb{R}^+ \times I''\times C\big)$, where $M'_3=I'\times C$ and $M''_3 = I''\times C$. This is illustrated in Fig.~\ref{fig:heegaardsplit} (right), where taking $C \to 0$, we indeed have $\mathbb{R}^+ \times I'$ and $\mathbb{R}^+ \times I''$.

\subsection{A Vafa-Witten Version of the Atiyah-Floer Correspondence} \label{subsec: VW AF Correspondence}

If $C \to 0$, we end up with an open $A$-model in complex structure $I$ on $\mathbb{R}^+\times I'$ and $\mathbb{R}^+\times I''$, respectively, with target space $\mathcal{M}^G_{\text{Higgs}}(C)$.
Because we have an $A$-model in complex structure $I$, the admissible branes are those of type $(A, *, *)$ 
Specifically, we need an $(A, *, *)$-brane in $\mathcal{M}_{\text{Higgs}}^G(C)$ that 
 corresponding to a Higgs pair on $C$ that can be extended to flat complex connections $\mathcal A$ on $M^{',''}_3$ 

Such an $(A, *, *)$-brane has indeed been obtained in~\cite{K08}.\footnote{The 4d theory considered in \cite{K08} is not the VW but the GL theory of \cite{KW06}, albeit with parameter $t=0$. However, both these 4d theories descend to the same 2d $A$-model with target $\mathcal{M}^G_{\text {Higgs}}(C)$ after dimensional reduction on $C$, and since our $A$-branes of interest are $A$-model objects within $\mathcal{M}^G_{\text {Higgs}}(C)$, the arguments used and examples stated in \cite{K08} are applicable here.\label{foot:usage of Kap}} It is an $(A,B,A)$-brane $\alpha_{M^{',''}_3}$, that is simultaneously an $A$-brane in $\mathcal{M}^G_{\text {Higgs}}(C)$ and an $A$-brane in ${\mathcal M}^G_H(C)$ in complex structure $K$, i.e., ${\mathcal M}^{G_\mathbb{C}}_{\text{flat}}(C)$, the moduli space of flat $G_\mathbb{C}$-connections on $C$, where it corresponds to flat connections that can be extended to $M_3^{',''}$. It is middle-dimensional, and is therefore a Lagrangian brane. Let us henceforth denote this brane as $L$.  

\begin{figure}
    \centering
    \includegraphics[width=0.95\textwidth]{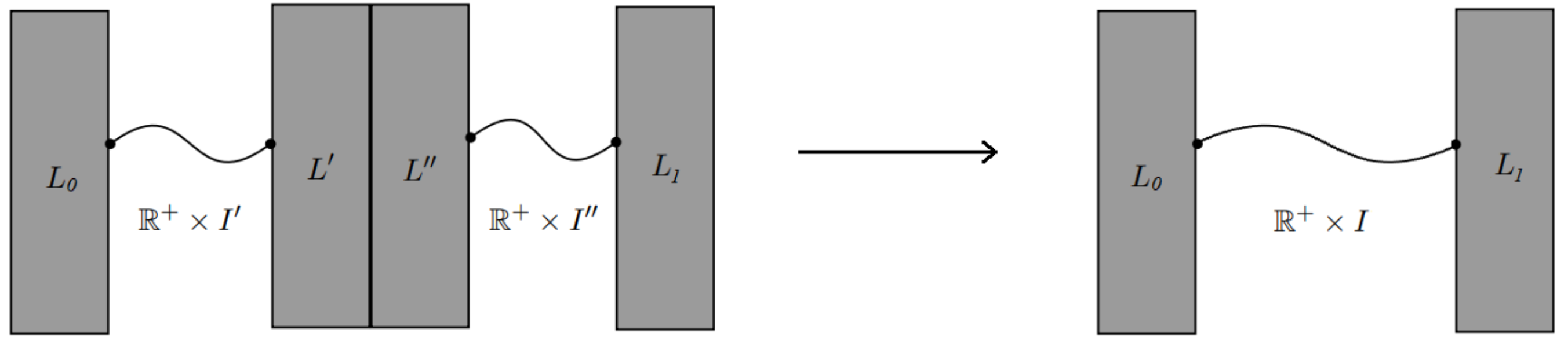} 
    \caption{Identifying $L'$ and $L''$  and gluing them together to form a single open string.}
    \label{fig:gluebranes}
\end{figure}

 Now, with two split pieces $M_4^{',''}$, when $C \to 0$, we have two strings, each ending on pairs of Lagrangian branes $(L_0, L')$ and $(L'', L_1)$ (see Fig.~\ref{fig:gluebranes}.)
We then glue the open worldsheets together along their common boundary $L^{',''}$, giving us a single $A$-model, with a single string extending from $L_0$ to $L_1$, which is equivalent to gluing $M^{',''}_4$ along $C \times \mathbb{R}^+$. (see Fig.~\ref{fig:gluebranes} again.)
As before, one can recast the $A$-model here as an SQM model, where $\mathbb{R}^+$ is `time', and the target space is $\mathscr P (L_0, L_1)$, the space of smooth trajectories from $L_0$ to $L_1$ (arising from the interval $I$ that connects them).  
The BPS equations for this $A$-model are \eqref{2dbpsfull}, i.e.,  holomorphic maps from the worldsheet to the target space. 
They can be written as a gradient flow equation on the worldsheet
\begin{equation}\label{worldsheetgradientflow}
    \frac{\partial Z^l}{\partial t} + i \frac{\partial Z^l}{\partial s} = 0,
\end{equation}
where we have used real coordinates $t$ and $s$ (for $z=t+is$), and here, $Z^l = X^l + Y^l$. 
The fixed critical points of the underlying potential of the SQM model that contribute to the partition function are defined by $  \dot{Z^l}  = {\partial Z^l / \partial s} = 0$. Since `$s$' is the spatial coordinate of $I$, it would mean that the fixed critical points just correspond to fixed stationary trajectories in $\mathscr P (L_0, L_1)$, i.e., the intersection points of $L_0$ and $L_1$. 
Thus, the partition function of the $A$-model, which, from the SQM model perspective, is given by an algebraic count of the fixed critical points of its underlying potential, will be an algebraic count of the intersection points of $L_0$ and $L_1$, where there are flow lines between the intersection points that obey \eqref{worldsheetgradientflow}. These flow lines correspond to holomorphic Whitney disks. 

From this description of the partition function, we have physically realized the Lagrangian Floer homology first defined in~\cite{F88a}, where the intersection points of $L_0$
and $L_1$ actually generate the chains of the Lagrangian Floer complex, and the Floer differential, which counts the
number of holomorphic Whitney disks, can be interpreted as the outgoing flow lines at each intersection point of $L_0$ and $L_1$ which number would be the degree of the corresponding chain in the complex. 

Specifically, let $(L_0\cap L_1)_i^{n_i}$ denote the $i^{\text{th}}$ point of the intersection $L_0\cap L_1$ where there are $n_i$ outgoing flow lines, whence we can identify
\begin{equation}
 (L_0\cap L_1)_i^{n_i} \in  \text{HF}^{\text{Lagr}}_{n_i}\big(\mathcal{M}^G_{\text{Higgs}}(C), L_0, L_1\big), 
\end{equation}
where $\text{HF}^{\text{Lagr}}_{n_i} \big(\mathcal{M}^G_{\text{Higgs}}(C), L_0, L_1 \big)$ is the Lagrangian Floer homology of $(L_0, L_1)$ on $\mathcal{M}^G_{\text{Higgs}}(C)$ of degree $n_i$. Then, the partition function of the $A$-model will be given by
\begin{equation} \label{ZA = HF^lag}
    \mathcal{Z}_{A,L}\big(\tau, \mathcal{M}^G_{\text{Higgs}}(C)\big) = \sum_i \text{HF}^{\text{Lagr}}_{n_i}\big(\mathcal{M}^G_{\text{Higgs}}(C), L_0, L_1, \tau\big),
\end{equation}
A $\tau$-dependency appears here because of a $\tau$-dependent term in the $A$-model action.

Since the underlying boundary VW theory on $M_4 = M_3 \times \mathbb{R}^+$ is topological, we will have the following equivalence of partition functions:
\begin{equation} \label{Zvw = Z_AL}
      \mathcal{Z}_{\text{VW},M_4}(\tau, G) = \mathcal{Z}_{A,L}\big(\tau, \mathcal{M}^G_{\text{Higgs}}(C)\big), 
\end{equation}
which, from \eqref{4d3dpartitionfinal} and \eqref{ZA = HF^lag}, means that 
\begin{equation}\label{atconj2}
    {\sum_k \text{HF}_{d_k}^{\text{VW}}(M_3, G, \tau)= \sum_{i}\text{HF}_{n_i}^{\text{Lagr}}\big(\mathcal{M}^G_{\text{Higgs}}(C), L_0, L_1, \tau\big).}
\end{equation}
The gradings `$d_k$' and `$n_i$' in \eqref{atconj2} match. To understand this, recall that the VW flow lines between fixed critical points in $\frak A$ are non-fixed solutions to the VW equations \eqref{4dbps0} on $M_3 \times \mathbb R^+$. Also, in $\S$\ref{4d bps to 2d bps}, it was shown that the VW equations descend to the worldsheet instanton equations \eqref{2dbpsfull} defining holomorphic maps from the worldsheet to $\mathcal{M}^G_{\text{Higgs}}(C)$, the non-fixed solutions to which are the flow lines between fixed critical points in $\mathscr P(L_0,L_1)$. Thus, there is a one-to-one correspondence between the flow lines that define $\text{HF}_{*}^{\text{VW}}$ and underlie the LHS of \eqref{atconj2}, and the flow lines that define $\text{HF}_{*}^{\text{Lagr}}$ and underlie the RHS of \eqref{atconj2}. Moreover,  `$k$' and `$i$' obviously match, too.

We thus have a degree-by-degree isomorphism of the VW Floer homology and the Lagrangian Floer homology, whence we would have a Vafa-Witten Atiyah-Floer correspondence 
\begin{equation}\label{atconj}
   \text{HF}_{*}^{\text{VW}}(M_3, G, \tau) \cong \text{HF}_{*}^{\text{Lagr}}\big(\mathcal{M}^G_{\text{Higgs}}(C), L_0, L_1, \tau\big).
\end{equation}

\subsection{A Physical Proof and Generalization of a Conjecture by Abouzaid-Manolescu about the Hypercohomology of a Perverse Sheaf of Vanishing Cycles}

A hypercohomology $\text{HP}^*(M_3)$ was constructed by Abouzaid-Manolescu in \cite{AM20}, where it was conjectured to be isomorphic to instanton Floer homology assigned to $M_3$ for the complex gauge group $SL(2,\mathbb{C})$.  
Its construction  was via a Heegaard split of $M_3=M'_3 \cup_{C}M''_3$ along $C$ of genus $g$, and the intersection of the two associated Lagrangians in the moduli space $X_{\text{irr}}(C)$  of irreducible flat $SL(2, \mathbb{C})$-connections on $C$ (that represent solutions extendable to $M'_3$ and $M''_3$, respectively), to which one can associate a perverse sheaf of vanishing cycles. $\text{HP}^*(M_3)$ is then the hypercohomology of this perverse sheaf of vanishing cycles in $X_{\text{irr}}(M_3)$, where it is an invariant of $M_3$ independent of the Heegaard split. 

Based on the mathematical construction of $\text{HP}^*(M_3)$ described above, it would mean that a physical realization of (the dual of) $\text{HP}^*(M_3)$ ought to be via an open $A$-model with Lagrangian branes $L_0$ and $L_1$ in the target $X_{\text{irr}}(C)$, where the observables contributing to the partition function can be interpreted as classes in the Lagrangian Floer homology $ \text{HF}_{*}^{\text{Lagr}}\big(X_{\text{irr}}(C), L_0, L_1, \tau\big)$.

First, note that there is an isomorphism between $\text{HF}_{*}^{\text{Lagr}}$ and the homology of Lagrangian submanifolds in $X_{\text{irr}}(C)$~\cite[Theorem 11]{P18}, i.e., 
\begin{equation}
    \text{HF}_{*}^{\text{Lagr}}\big(X_{\text{irr}}(C), L_0, L_1, \tau\big)\cong\text{H}_*(L, \mathbb{Z}_2)_{\otimes\mathbb{Z}_2}\Lambda,
\end{equation}
where $\Lambda$ is a scalar function over $\mathbb{Z}_2$, called the Novikov field, and $L$ on the RHS can be taken as either $L_0$ or $L_1$. The homology cycles of the  Lagrangian (i.e., middle-dimensional) submanifolds of $X_{\text{irr}}(C)$ have a maximum dimension of $\frac{1}{2}\text{dim}(X_{\text{irr}}(C))$, where $\frac{1}{2}\text{dim}(X_{\text{irr}}(C))=2(3g-3)$.\footnote{It is a fact that $\text{dim}(X_{\text{irr}}(C))$ is given by $4 (N^2-1)(g-1)$ for $G_{\mathbb{C}} = SL(N, \mathbb{C})$, where $g$ is the genus of $C$.} Including the zero-cycle, the grading of $\text{H}_*(L, \mathbb{Z}_2)_{\otimes\mathbb{Z}_2}\Lambda$ and therefore $\text{HF}_{*}^{\text{Lagr}}\big(X_{\text{irr}}(C), L_0, L_1, \tau\big)$, goes as  $0, 1, \dots, 2(3g-3)$. 


 Second, note that in~\cite[Theorem 1.8]{AM20}, it was computed that $\text{HP}^k$ is nonvanishing only if $-3g+3\leq k \leq 3g-3$. In other words, the grading of $\text{HP}^*$ goes as $-(3g-3), \dots, 0, \dots, (3g-3)$. 
 
 These two observations then mean that there is a one-to-one correspondence between the gradings of $\text{HP}^*(M_3)$  and $\text{HF}_{*}^{\text{Lagr}}$. Moreover, the generators of $\text{HP}^*$  and $\text{HF}_{*}^{\text{Lagr}}$ both originate from the intersection points of $L_0$ and $L_1$ in $X_{\text{irr}}(C)$. Hence, we can identify $\text{HP}^*$  with (the dual of) $\text{HF}_{*}^{\text{Lagr}}$, i.e., 
\begin{equation} \label{HP - HF^lag}
    \text{HP}^*(M_3) \cong \text{HF}_{*}^{\text{Lagr}}\big(X_{\text{irr}}(C), L_0, L_1, \tau\big)
 \end{equation} 
This agrees with~\cite[Remark 6.15]{BB12}. 

Notice from the Morse functional \eqref{complex CS fn} and the gradient flow equation \eqref{flatcomplex - final} that the definition of $\text{HF}_*^{\text{VW}}$ coincides with the definition of the instanton Floer homology in~\cite{F88b}, albeit for a \emph{complex} gauge group $G_{\mathbb{C}}$. This means that we can also express the LHS of \eqref{atconj} as $\text{HF}_*^{\text{Inst}}\big(M_3, G_{\mathbb{C}}, \tau \big)$, the instanton Floer homology of $G_{\mathbb{C}}$ assigned to $M_3$. 
Also, recall that the Lagrangian branes $L_0$ and $L_1$ on the RHS of \eqref{atconj} are $(A,B,A)$-branes, i.e., they can also be interpreted as Lagrangian branes in ${\mathcal M}^G_H(C)$ in complex structure $K$, or equivalently, ${\mathcal M}^{G_{\mathbb{C}}}_{\text{flat}}(C)$, the moduli space of irreducible flat $G_{\mathbb{C}}$-connections on $C$.

These two points then mean that we can also write \eqref{atconj} as
\begin{equation}\label{HF^inst = HF^lag(Mflat)}
   \text{HF}_{*}^{\text{inst}}(M_3, G_{\mathbb{C}}, \tau) \cong 
 \text{HF}_{*}^{\text{Lagr}}\big({\mathcal M}^{G_{\mathbb{C}}}_{\text{flat}}(C), L_0, L_1, \tau\big)
\end{equation}
In other words, the VW Atiyah-Floer correspondence in \eqref{atconj} can also be interpreted as an Atiyah-Floer correspondence for $G_{\mathbb{C}}$-instantons. 
 It is now clear from \eqref{HF^inst = HF^lag(Mflat)} and \eqref{HP - HF^lag}, that for $G_\mathbb C = SL(2, \mathbb{C})$, we have
\begin{equation} \label{AB-Mano conj}
  \text{HP}^*(M_3) \cong \text{HF}_{*}^{\text{inst}}(M_3, SL(2, \mathbb{C}), \tau) 
\end{equation}
for complex constant $\tau$. This is exactly the conjecture by Abouzaid-Manolescu about $\text{HP}^*(M_3)$ in~\cite{AM20}! 
This agrees with their expectations in~\cite[sect.~9.2]{AM20} that $\text{HP}^*(M_3)$ ought to be part of 3+1 dimensional TQFT based on the VW equations.

It was argued in~\cite[sect.~9.1]{AM20} that the construction of $\text{HP}^*(M_3)$ can be generalized to $SL(N, \mathbb{C})$. Indeed, notice that \eqref{HF^inst = HF^lag(Mflat)} implies that there ought to be a $G_\mathbb{C}$ generalization of the Abouzaid-Manolescu conjecture in \eqref{AB-Mano conj} to 
\begin{equation} \label{AB-Mano conj generalized}
  \text{HP}^*(M_3, G_\mathbb{C}) \cong \text{HF}_{*}^{\text{inst}}(M_3, G_\mathbb{C}, \tau) 
\end{equation}
where the hypercohomology $\text{HP}^*(M_3, G_\mathbb C)$ of the perverse sheaf of vanishing cycles in ${\mathcal M}^{G_\mathbb{C}}_{\text{flat}}(M_3)$ is such that
\begin{equation} \label{HP - HF^lag general}
   \text{HP}^*(M_3, G_\mathbb{C}) \cong \text{HF}_{*}^{\text{Lagr}}\big({\mathcal M}^{G_\mathbb{C}}_{\text{flat}}(C), L_0, L_1, \tau\big)
 \end{equation} 
which again agrees with~\cite[Remark 6.15]{BB12}. 

\section{Langlands Duality of Vafa-Witten Invariants, Gromov-Witten invariants, Floer Homologies and the Abouzaid-Manolescu Hypercohomology }\label{sec:langlands duality} 

It is known that $\mathcal{N}=4$ supersymmetric Yang-Mills theories has a $SL(2, \mathbb{Z})$ symmetry, with $S$- and $T$-duality, as mentioned in $\S$\ref{vwgeneral}. In particular, the theory with complex coupling $\tau$ and gauge group $G$, is $S$-dual to a theory with complex coupling $-\frac{1}{n_{\mathfrak g}\tau}$ and Langlands dual gauge group $^L G$, 
i.e., we have, up to a possible phase factor of modular weights that is just a constant, a duality of VW partition functions
\begin{equation}\label{vwdual}
   \mathcal{Z}_{\text{VW},M_4}(\tau, G) \longleftrightarrow \mathcal{Z}_{\text{VW},M_4}\Big(-\frac{1}{n_{\mathfrak{g}}\tau},\, ^LG \Big)
\end{equation}
In other words, we have a Langlands duality of VW invariants of $M_4$, given by \eqref{vwdual}.


\subsection{Langlands Duality of Gromov-Witten Invariants}

Note that if $M_4 = \Sigma \times C$, from \eqref{vwdual} and \eqref{VW=GW}, 4d $S$-duality would mean that we have the 2d duality 
\begin{equation}\label{ZGW=ZGW}
    \mathcal{Z}_{\text{GW},\Sigma}\big(\tau, \mathcal{M}^{G}_{\text{Higgs}}(C)\big) 	\longleftrightarrow \mathcal{Z}_{\text{GW},\Sigma}\Big(-\frac{1}{n_{\mathfrak{g}}\tau}, \,\mathcal{M}^{^LG}_{\text{Higgs}}(C)\Big)
\end{equation}
 where $\mathcal{M}^{G}_{\text{Higgs}}$ and $\mathcal{M}^{^LG}_{\text{Higgs}}$ are mirror manifolds. 
 In other words, we have a Langlands duality of GW invariants 
 that can be interpreted as a mirror symmetry of Higgs bundles, given by \eqref{ZGW=ZGW}.  


\subsection{Langlands Duality of Vafa-Witten Floer Homology}

If $M_4 = M_3 \times \mathbb{R}^+$, from \eqref{4d3dpartitionfinal} and \eqref{vwdual}, we have the duality 
\begin{equation}\label{ZFL to ZFLG}
  \mathcal{Z}^{\text{Floer}}_{\text{VW},M_3}(\tau, G) \longleftrightarrow \mathcal{Z}^{\text{Floer}}_{\text{VW},M_3}\Big(-\frac{1}{n_{\mathfrak{g}}\tau},\, ^LG \Big).
\end{equation}
In turn, from \eqref{4d3dpartitionfinal}, this means that we have the duality
\begin{equation}\label{HFVWG to HFVWLG}
  \text{HF}^{\text{VW}}_*(M_3, G, \tau) \longleftrightarrow \text{HF}^{\text{VW}}_*(M_3, {^LG}, -1/n_{\mathfrak{g}}\tau)
\end{equation}

In other words, we have a Langlands duality of VW Floer homologies assigned to $M_3$, given by \eqref{HFVWG to HFVWLG}.

\subsection{Langlands Duality of Lagrangian Floer Homology}

From \eqref{ZFL to ZFLG} and \eqref{Zvw = Z_AL}, we have the duality
\begin{equation}\label{ZAL to ZAL}
   \mathcal{Z}_{A,L}\big(\tau, \mathcal{M}^G_{\text{Higgs}}(C)\big)
 \longleftrightarrow \mathcal{Z}_{A,L}\Big(-\frac{1}{n_{\mathfrak{g}}\tau},\, ^LG \Big).
\end{equation}
Then, from the RHS of the VW Atiyah-Floer correspondence in \eqref{atconj}, which defines the state spectrum of $\mathcal{Z}_{A,L}$, we have the duality
\begin{equation} \label{HFL to HFL}
\text{HF}_{*}^{\text{Lagr}}\big(\mathcal{M}^G_{\text{Higgs}}(C), L_0, L_1, \tau\big) \longleftrightarrow \text{HF}_{*}^{\text{Lagr}}\big(\mathcal{M}^{^LG}_{\text{Higgs}}(C), L_0, L_1, -1/n_{\mathfrak{g}}\tau\big)
\end{equation}

In other words, we have a Langlands duality of Lagrangian Floer homologies of Higgs bundles, given by \eqref{HFL to HFL}.

\subsection{Langlands Duality of the Abouzaid-Manolescu Hypercohomology}

From \eqref{AB-Mano conj generalized}, the fact that its RHS can be identified with $\text{HF}^{\text{VW}}_*(M_3, G, \tau)$, and the relation \eqref{HFVWG to HFVWLG}, we have the duality
\begin{equation} \label{Abou-Mano Langlands}
   \text{HP}^*(M_3, G_\mathbb{C}, \tau) \longleftrightarrow \text{HP}^*(M_3, ^LG_\mathbb{C}, - 1/n_\mathfrak g\tau) 
\end{equation}

In other words, we have a Langlands duality of the Abouzaid-Manolescu hypercohomologies of a perverse sheaf of vanishing cycles in the moduli space of irreducible flat  complex connections on $M_3$, given by \eqref{Abou-Mano Langlands}. 

\section{A Quantum and Classical Geometric Langlands Correspondence}

If we let $M_4 = I \times \mathbb{R}^+ \times C$ with $C \to 0$, $S$-duality gives a homological mirror symmetry of the category of $A$-branes.
This implies a homological mirror symmetry of the $\tau$-dependent category of $A$-branes:
\begin{equation}\label{dualcata}
   \text{Cat}_{\text{$A$-branes}}\big(\tau, \mathcal{M}^{G}_{\text{Higgs}}(C)\big) \longleftrightarrow
    \text{Cat}_{\text{$A$-branes}}\Big(-\frac{1}{n_{\mathfrak{g}}\tau}, \,\mathcal{M}^{^LG}_{\text{Higgs}}(C)\Big) 
\end{equation}
where $\mathcal{M}^{G}_{\text{Higgs}}$ and $\mathcal{M}^{^LG}_{\text{Higgs}}$ are mirror manifolds. 

For $\theta=0$ (Re$(\tau)=0$), the category of $\tau$-dependent $A$-branes can be identified with a category of twisted $D$-modules on $\text{Bun}_{G_{\mathbb{C}}}(C)$ with parameter $q$. Thus, this mirror symmetry would mean that we have 
\begin{equation}\label{dualdmod}
    {\mathcal D}^{\textbf{c}}_{-h^\vee}\text{-mod}\big(q, {\text{Bun}_{G_{\mathbb{C}}}}\big) \longleftrightarrow {\mathcal D}^{\textbf{c}}_{-^Lh^\vee}\text{-mod}\Big(-\frac{1}{n_{\mathfrak{g}}q}, \,{\text{Bun}_{{^LG}_{\mathbb{C}}}}\Big)
\end{equation}
This is a quantum geometric Langlands correspondence for $G_{\mathbb{C}}$ with complex curve $C$ and purely imaginary parameter $q$ \cite[eqn.~(6.4)]{F05}. 

On the other hand, in the zero-coupling, `classical' limit of the 4d theory in $G$ where $\text{Im}(\tau) \to \infty$, we have $q \to \infty$. In this limit, the LHS of \eqref{dualdmod} can be identified with the category $\text{Cat}_{\text{coh}}\big (\mathcal{M}_{\text{flat}}^{G_{\mathbb{C}}}(C) \big )$ of coherent sheaves on $\mathcal{M}_{\text{flat}}^{G_{\mathbb{C}}}(C)$~\cite{F05}.     
This `classical' limit corresponds to the `ultra-quantum' limit of the $S$-dual 4d theory in $^LG$, where $^Lq = -\frac{1}{n_{\mathfrak{g}}q} \to 0$. In this limit, the RHS of \eqref{dualdmod} can be identified with the category ${\mathcal D}^{\textbf{c}}_{-{^Lh}^\vee}\text{-mod}\big(0, {\text{Bun}_{{^LG}_{\mathbb{C}}}}\big) $ of critically-twisted $D$-modules on $\text{Bun}_{{^LG}_{\mathbb{C}}}(C)$, giving us
\begin{equation}\label{classical GL}
    \text{Cat}_{\text{coh}}\big (\mathcal{M}_{\text{flat}}^{G_{\mathbb{C}}}(C) \big ) \longleftrightarrow {\mathcal D}^{\textbf{c}}_{-^Lh^\vee}\text{-mod}\Big(0, \,{\text{Bun}_{{^LG}_{\mathbb{C}}}}\Big)
    \end{equation}
This is a classical geometric Langlands correspondence for $G_{\mathbb{C}}$ with complex curve $C$ \cite[eqn.~(6.4)]{F05}.

\section{Categorification and a Novel Web of Mathematical Relations}\label{sec:web of dualities}

The mathematical procedure  of categorification is realized in our physical framework, where the VW invariant is a number, the VW Floer homology is a vector (space), and the $A$-branes span a category of objects.
Categorification can be physically understood as flattening a direction and then ending it on a boundary or boundaries. Explicitly in our case, the first step of categorification involves flattening a direction in $M_4$ and then ending it on an $M_3$ boundary, while the second step involves flattening a direction in $M_3$ and then ending it on two $C$ boundaries. Therefore, one can also understand the procedure of categorifying as computing relative invariants\footnote{A relative invariant is an invariant of an open manifold which was originally defined for a closed manifold.} -- computing the relative invariant of ${\mathcal Z}_{\text {VW}}$ give us ${\text {HF}}^{\text {VW}}_*$, and further computing the relative invariant of ${\text {HF}}^{\text {VW}}_*$ gives us $\text{Cat}_{A\text{-branes}}$. 

One could continue to further categorify the VW invariant of $M_4$ by flattening a direction along $C$ and ending it on $S^1$ boundaries, i.e., let $C = I' \times S^1$. This should give us a 2-category, 2-Cat, consisting of objects, morphisms between these objects, and 2-morphisms between these morphisms.
We thus have\footnote{This perspective of categorifying topological invariants by successively introducing boundaries to the manifold was first pointed out in \cite{G09}.}
\begin{equation}\label{category of objects- extend}
    \begin{split}
        \text{VW theory on } M_4   \quad &\leadsto  \quad \text{number} \quad \qquad \mathcal{Z}_{\text{VW}} \\
        \text{VW theory on } \mathbb{R}^+ \times M_3 \quad &\leadsto \quad  \text{vector }\qquad\quad \text {HF}^{\text {VW}}_*\\
        \text{VW theory on } \mathbb{R}^+ \times I \times C \quad &\leadsto \quad  \text{1-category } \quad\text{Cat}_{A\text{-branes}}\\
        \text{VW theory on } \mathbb{R}^+ \times I \times I' \times S^1 \quad &\leadsto  \quad \text{2-category } \quad \text{2-Cat}\\
        \text{VW theory on } \mathbb{R}^+ \times I \times I' \times [0,1] \quad &\leadsto  \quad \text{3-category } \quad\text{3-Cat}.
    \end{split}
\end{equation}
As we go down the list, the categories get assigned to $M_3, C, \dots$, and are determined by the category of boundaries of the effective 1d, 2d, ... theory on $\mathbb R^+, \mathbb R^+ \times I, \dots$. Therefore, the 2-category will be determined by the category of 2d boundaries of the 3d theory on $\mathbb R^+ \times I \times I'$ given by VW theory compactified on $S^1$, that is assigned to $S^1$. These are surface defects that can be interpreted as objects; loop defects on the surface running around $I \times I'$ can be interpreted as morphisms between these objects; while opposing pairs of point defects on the loops can be interpreted as 2-morphisms between these morphisms. 
 Also, note that the 3d TQFT in question is a 3d gauged $A$-model described in \cite[sect. 7]{KSV10},\footnote{In \cite[sect. 7]{KSV10}, the GL theory at $t=0$ was considered, but it was shown in \cite[sect. 5.2-5.3]{S13} that this theory compactified on $S^1$ is the same as VW theory compactified on $S^1$. Hence, their results are applicable to us.} and for abelian $G$ and $\text{Re}(\tau)=0$, the 2-category of surface defects have been explicitly determined in $\it{loc.\,cit.}$ to be the 2-category $\text{2-Cat}_{\text{mod-cat}} \big({\text{FF-cat}}(T^2)\big)$ of module categories over the Fukaya-Floer category of $T^2$.\footnote{The 3d gauged $A$-model has a gauge and matter sector, where each sector can either have Dirichlet (D) or Neumann (N) boundary conditions. We have stated the result for the DD case, as this choice of boundary conditions allows us to describe the situation where line defects lie along the surface defects, which is the one relevant to us.}

4d $S$-duality also gives us a Langlands duality of the 2-category 2-Cat.
According to \cite[sect. 7.4.1]{KSV10}, 4d $S$-duality, which maps abelian $G$ to its Langlands dual that is itself, will transform the symplectic area $\mathcal A$ of $T^2$ as
\begin{equation}
    {\mathcal A} \to {^L{\mathcal A}} = {\frac{4 \pi^2}  {\mathcal A}},
\end{equation}
where ${^L{\mathcal A}}$ is the symplectic area of a torus $^LT^2$ that can be obtained from $T^2$ by inverting the radii of its two circles from $R \to {\alpha'/ R}$ for some constant $\alpha'$. In other words, $^LT^2$ is the $T$-dual torus to $T^2$, and FF-cat($T^2$), which is realized by a 2d open $A$-model with target $T^2$, will be invariant under $T$-duality of the target, i.e., FF-cat($T^2$) $\cong$ FF-cat($^LT^2$). Thus, we have
\begin{equation} \label{2- Cat to 2-Cat}
  \text{2-Cat}_{\text{mod-cat}} \big({\text{FF-cat}}(T^2)\big) \longleftrightarrow \text{2-Cat}_{\text{mod-cat}} \big({\text{FF-cat}}(^LT^2)\big) .
\end{equation}

The last step to further categorify the VW invariant of $M_4$ is to flatten $S^1$, ending it on point boundaries, i.e., let $S^1 = [0,1]$. This should give us a 3-category, 3-Cat, consisting of objects, morphisms between these objects, 2-morphisms between these morphisms, and 3-morphisms between these 2-morphisms, giving us a 3-category of 3d boundary conditions of  VW theory along $\mathbb R^+ \times I \times I'$ which is assigned to a point. These 3d boundary conditions can be realized by domain walls. 

From the duality relations \eqref{vwdual}, \eqref{ZGW=ZGW}, \eqref{HFVWG to HFVWLG}, \eqref{HFL to HFL}, the correspondences \eqref{dualcata}, \eqref{dualdmod}, \eqref{classical GL}, and the identifications \eqref{VW=GW}, \eqref{4d3dpartition}, \eqref{atconj}, we will get Fig.~\ref{fig:equimath - cat} below.

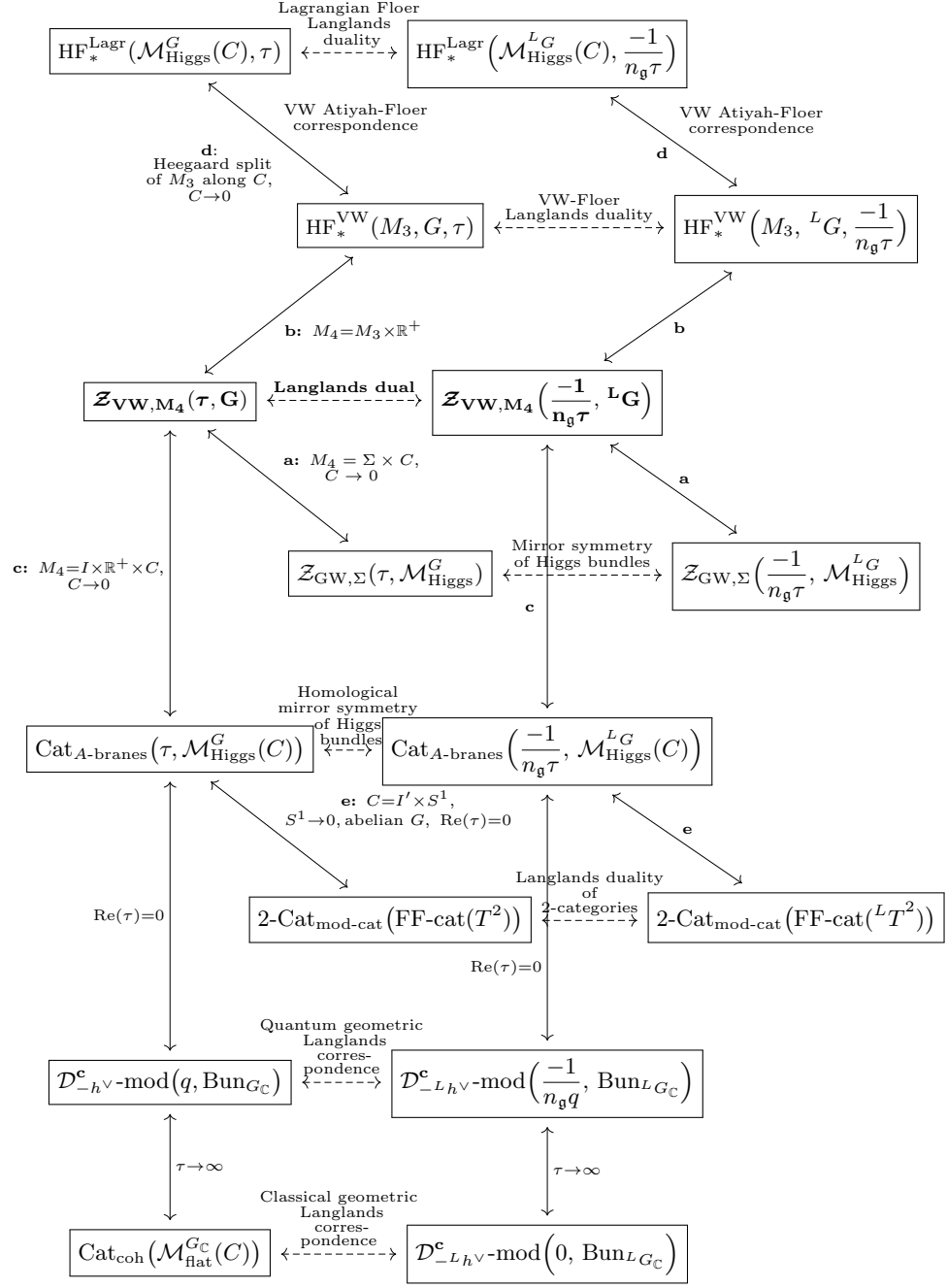
\begin{figure}\small
 \centering
\begin{tikzcd}[row sep=35, column sep=-33]
\boxed{\footnotesize{\text{HF}^{\text{Lagr}}_\ast (\mathcal{M}^G_{\text{Higgs}}(C), \tau)}} 
\arrow[rrr, leftrightarrow, dashed, "\substack{\text{Lagrangian Floer}\\\text{Langlands}\\\text{ duality}}"]
\arrow[dr, leftrightarrow,  " \substack{\text{VW Atiyah-Floer}\\\text{correspondence}}", " \substack{\textbf{d}:\\\text{ Heegaard split}\\ \text{of $M_3$ along $C$,}\\C\to 0}"']
&&&\boxed{\footnotesize{\text{HF}^{\text{Lagr}}_\ast \Big(\mathcal{M}^{^LG}_{\text{Higgs}}(C),  \frac{-1}{n_{\mathfrak{g}}\tau}\Big)}} 
\arrow[dr, leftrightarrow,  " \substack{\text{VW Atiyah-Floer}\\\text{correspondence}}", "\textbf{d}"']
\\
&\boxed{\footnotesize{\text{HF}^{\text{VW}}_\ast(M_3, G, \tau)}}  
\arrow[rrr, leftrightarrow, dashed,"\substack{\text{VW-Floer}\\\text{Langlands duality}}"]
\arrow[dl, leftrightarrow, "\textbf{b: }M_4=M_3 \times \mathbb{R}^+"]
&&&\boxed{\footnotesize{\text{HF}^{\text{VW}}_\ast \Big(M_3,\, ^LG, \frac{-1}{n_{\mathfrak{g}}\tau}\Big)}} 
\arrow[dl, leftrightarrow, "\textbf{b }"]\\
\boxed{\footnotesize{\bf{\boldsymbol{\mathcal{Z}}_{\textbf{VW},M_4}(\boldsymbol{\tau}, G)} }}
\arrow[rrr, leftrightarrow, dashed, "\textbf{Langlands dual}"]
\arrow[dr, leftrightarrow, "\substack{\textbf{a: }\text{$M_4=\Sigma \times C$,}\\ \text{$C\to 0$}}"]
&&&\boxed{\footnotesize{\bf{\boldsymbol{\mathcal{Z}}_{\textbf{VW},M_4}\Big(\frac{-1}{n_{\mathfrak{g}}\boldsymbol{\tau}},\, ^LG \Big)} }}
\arrow[dr, leftrightarrow, "\textbf{a }"]\\
&\boxed{\footnotesize{\mathcal{Z}_{\text{GW},\Sigma}(\tau, \mathcal{M}^G_{\text{Higgs}})}}
\arrow[rrr, leftrightarrow, dashed, "\substack{\text{Mirror symmetry}\\\text{of Higgs bundles}}", crossing over]
&&&\boxed{\footnotesize{\mathcal{Z}_{\text{GW},\Sigma}\Big(\frac{-1}{n_{\mathfrak{g}}\tau},\,\mathcal{M}^{^LG}_{\text{Higgs}}\Big)}}  \\
\boxed{\text{Cat}_{\text{$A$-branes}}\big(\tau, \mathcal{M}^{G}_{\text{Higgs}}(C)\big)}
\arrow[dd,leftrightarrow,  "\text{Re}(\tau) = 0"']
\arrow[dr, leftrightarrow, " \substack{\textbf{e: } C = I' \times S^1,\\ S^1 \to 0,\, \text{abelian $G$, }\,\text{Re}(\tau) = 0}"]
\arrow[uu, leftrightarrow,  " \substack{{\textbf{c: }M_4=I\times\mathbb{R}^+\times C,}\\{ C\to 0}}"] 
\arrow[rrr, leftrightarrow, dashed, "\substack{\text{Homological}\\\text{ mirror symmetry }\\\text{of Higgs}\\\text{ bundles}}", crossing over]
&&&\boxed{\text{Cat}_{\text{$A$-branes}}\Big(\frac{-1}{n_{\mathfrak{g}}\tau}, \,\mathcal{M}^{^LG}_{\text{Higgs}}(C)\Big)}
\arrow[uu, leftrightarrow, "\substack{\\\\\\\\\\\textbf{c }}"]
\arrow[dd, leftrightarrow, "\substack{\\\\\\\\\\\\\\\\\text{Re}(\tau) = 0}"']
\arrow[dr, leftrightarrow, " \textbf{e}"]\\
&\boxed{\text{2-Cat}_{\footnotesize{\text{mod-cat}}} \big({\text{FF-cat}}(T^2)\big)}
\arrow[rrr, leftrightarrow, dashed, "\substack{{\text{Langlands duality}}\\{\text{ of }}\\{\text{2-categories}}}"]
&&&\boxed{\text{2-Cat}_{\text{mod-cat}} \big({\text{FF-cat}}({^LT}^2)\big)}\\
\boxed{{\mathcal D}^{\textbf{c}}_{-h^\vee}\text{-mod}\big(q, {\text{Bun}_{G_{\mathbb{C}}}}\big)}
\arrow[rrr, leftrightarrow, dashed, "\substack{{\text{Quantum geometric}}\\{\text{ Langlands}}\\{\text{ corres-}}\\\text{pondence}}"]
\arrow[d, leftrightarrow, "\tau \to \infty"]
&&&\boxed{{\mathcal D}^{\textbf{c}}_{-^Lh^\vee}\text{-mod}\Big(\frac{-1}{n_{\mathfrak{g}}q}, \,{\text{Bun}_{^L{G_{\mathbb{C}}}}}\Big)}
\arrow[d, leftrightarrow, "\tau \to \infty"]\\
\boxed{\text{Cat}_{\text{coh}}\big (\mathcal{M}_{\text{flat}}^{G_{\mathbb{C}}}(C) \big )}
\arrow[rrr, leftrightarrow, dashed, "\substack{{\text{Classical geometric}}\\{\text{ Langlands}}\\{\text{ corres-}}\\\text{pondence}}"]
&&&\boxed{{\mathcal D}^{\textbf{c}}_{-^Lh^\vee}\text{-mod}\Big(0, \,{\text{Bun}_{{^LG}_{\mathbb{C}}}}\Big)}
\end{tikzcd}
\caption{A novel web of mathematical relations stemming from Vafa-Witten theory that also involves higher categories.}
\label{fig:equimath - cat}
\end{figure}


\bibliographystyle{amsalpha}

\end{document}